\documentclass[pdflatex,sn-nature, referee]{sn-jnl}
\usepackage{graphicx}
\usepackage{tabularray}
\usepackage{enumitem}
\usepackage[labelfont=bf]{caption}
\newcommand\m[1]{\begin{pmatrix}#1\end{pmatrix}} 
\hypersetup{
    colorlinks=true,
    allcolors=black
}
\usepackage{lineno}
\usepackage{multirow}
\usepackage{amsmath}
\usepackage{amssymb}
\usepackage{amsthm}
\usepackage{bbm}
\usepackage{mathrsfs}
\usepackage[title]{appendix}
\usepackage{xcolor}
\usepackage{textcomp}
\usepackage{manyfoot}
\usepackage{booktabs}
\usepackage{algorithm}
\usepackage{algorithmicx}
\usepackage{algpseudocode}
\usepackage{listings}
\usepackage{verbatim}

\theoremstyle{thmstyleone}%

\theoremstyle{thmstyletwo}%

\theoremstyle{thmstylethree}%

\begin{document}

\title[Article Title]{Spatial Generalization Tests for Machine Learning-based Weather Models to Assess Physical Consistency}

\author*[1,2]{\fnm{Maren} \sur{Höver}}\email{maren.hoever@physics.ox.ac.uk}

\author[1]{\fnm{Milan} \sur{Klöwer}}

\author[3]{\fnm{Christian} \sur{Schroeder de Witt}}

\author[1]{\fnm{Hannah M.} \sur{Christensen}}

\affil*[1]{\orgdiv{Atmospheric, Oceanic and Planetary Physics}, \orgname{University of Oxford}, 
\country{United Kingdom}}

\affil[2]{\orgdiv{Intelligent Earth UKRI Centre for Doctoral Training in AI for the Environment}, \orgname{University of Oxford}, 
\country{United Kingdom}}

\affil[3]{\orgdiv{Engineering Science}, \orgname{University of Oxford},
\country{United Kingdom}}

\abstract{Machine learning-based weather prediction is revolutionizing weather forecasting by learning from weather data in present-day climate.
However, generalization to other climates remains a major challenge.
With melting sea ice, land-use change, and increasing ocean temperatures, boundary conditions are changing.
Therefore, generalization in time depends on generalization in space.
Here, we present three test cases to evaluate whether machine learning-based weather and climate models generalize in space and apply them to GraphCast and NeuralGCM.
We reverse or rotate the planet in longitude or latitude under the model's coordinate system and adapt all boundary conditions and forcings accordingly.
Physics-based general circulation models simulate a rotated/reversed planet with only rounding errors, but GraphCast and NeuralGCM fail these tests.
The analyses furthermore revealed unphysical variable mappings based on correlation rather than causation.
We argue that machine learning-based climate models should be designed to pass generalization tests to prevent overfitting on present-day regional climate.}

\maketitle

\section{Introduction}\label{sec1}

Machine learning-based weather prediction (MLWP) is revolutionizing weather forecasting.
Starting with the graph neural network-based model by Keisler~\cite{keislerForecastingGlobalWeather2022}, both purely data-driven models, such as GraphCast~\citep{lamLearningSkillfulMediumrange2023}, Pangu-Weather \citep{biAccurateMediumrangeGlobal2023} or GenCast~\citep{priceProbabilisticWeatherForecasting2025}, and hybrid general circulation models, like NeuralGCM~\citep{kochkovNeuralGeneralCirculation2024}, are now beating established numerical weather prediction models in some forecast scores~\citep{Rasp2020weatherbench}.
NeuralGCM, for instance, demonstrated that learning from weather forecasts of up to 5 days can yield largely numerically stable climate simulations over decades.
Before that, fully data-driven models like GraphCast\citep{lamLearningSkillfulMediumrange2023} or FourCastNet \citep{pathakFourCastNetGlobalDatadriven2022a} were able to produce skillful forecasts rivaling those of the best physics-based weather models.
However, they usually converge towards climatology within weeks (“blurring”, \cite{mathieuDeepMultiscaleVideo2016, lamLearningSkillfulMediumrange2023}), making them unsuitable for simulations considerably longer than they are trained for.
Other machine learning-based weather models suffer from instabilities during longer integrations, autoregressively diverging from any physical realism before hitting arithmetic overflow (“blow up", \cite{karlbauerAdvancingParsimoniousDeep2024b}).
Overall, machine learning-based weather models have the potential to provide more accurate weather predictions, particularly at a lower computational cost operationally, which enables the generation of larger ensembles~\citep{maheshHugeEnsemblesPart2025}.
However, during training, machine learning-based models continue to be associated with high computational cost.
Larger time steps of 6 or 12 hours, compared to minutes, mean they do not resolve sub-daily weather variation well.
For longer integrations, it remains to be tested whether these models have learned general physics, which is necessary for climate-scale application.

In contrast to weather prediction, \textit{climate} prediction based on machine learning is just emerging as a field \citep{watt-meyerACEFastSkillful2023, Ullrich2025, Henn2026}, with additional challenges arising:
Longer timescales, larger complexity added by ocean, land, ice, or chemical processes, as well as changing natural and anthropogenic boundary conditions.
Models like ACE~\citep{watt-meyerACEFastSkillful2023} emulate existing physics-based numerical climate models.
When machine learning-based \textit{weather} prediction models, trained on ERA5 data~\citep{hersbachERA5GlobalReanalysis2020}, are tested in warmer or colder climates than they were trained on, they frequently exhibit cold or warm biases, respectively~\citep{rackowRobustnessAIbasedWeather2024}.
Simultaneously, these models maintain comparable forecasting scores, providing evidence that forecast scores alone may not be a sufficient optimization metric.

While forecast skill metrics have been the primary tool for MLWP model evaluation~\citep{raspWeatherBench2Benchmark2024}, some studies have also evaluated their physical consistency. This includes assessments of their ability to reproduce atmospheric dynamics, such as Kelvin and Rossby waves \citep{jalanIntraseasonalEquatorialKelvin2025}, Tropical-Extratropical Teleconnections according to the Madden-Julian-Oscillation \citep{diaoAssessingMJOTropicalExtratropical2025}, and extreme weather events \citep{bano-medinaAreAIWeather2025, olivettiDatadrivenModelsBeat2024, mengDeepLearningAtmospheric2025, zhangPhysicsbasedModelsOutperform2026}, as well as geostrophic balance and power spectra \citep{bonavitaLimitationsCurrentMachine2024}, error-growth characteristics \citep{selzCanArtificialIntelligenceBased2023}, and the dynamical response to local perturbations \citep{hakimDynamicalTestsDeep2024}.

One key aspect of physical consistency in MLWP models that has not been investigated in detail is spatial generalization. The physical laws of the atmosphere are invariant in space and time. While variables and boundary conditions do depend on space and time, the governing equations determining the future evolution do not explicitly.

As such, we claim that for climate models to be trusted with out-of-distribution predictions, they need to demonstrate physical invariance to changing boundary conditions driven by climate change.
For example, a hard-coded learned representation of local atmospheric processes above ice is likely inaccurate under global warming when the ice boundary will have moved.
However, a model that learned invariant physical laws would be able to apply learned physics from previously ice-free regions to those where melting takes place.
As such, the model should explicitly depend on the presence or properties of ice rather than the coordinates where ice exists in the training data.
In this study, we formulate spatial generalization test cases for machine learning-based atmospheric models.
We argue that such tests are important to pass when building robust climate models that generalize to changing boundary conditions, such as melting ice and land-use change.
These tests are applied to the models GraphCast and NeuralGCM, and their performance is discussed.
We encourage developers to build machine learning-based models that pass our tests to prevent overfitting on regional present-day climate in the training data.

\begin{figure}[htbp]
    \noindent\includegraphics[width=\textwidth]{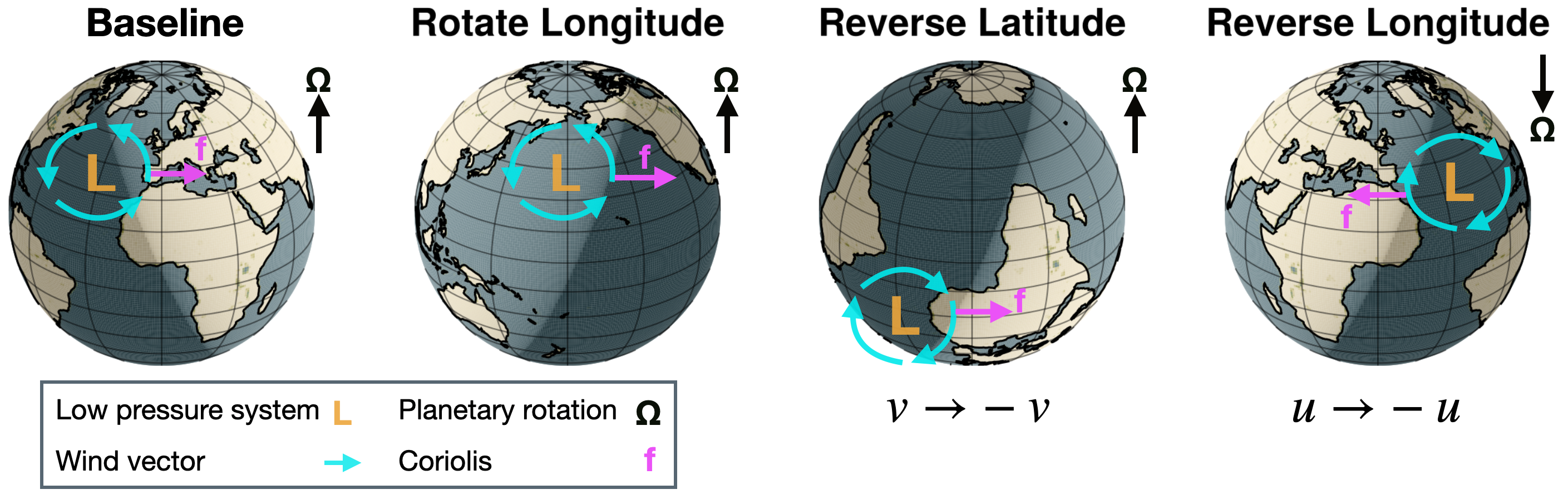}
    \caption{\textbf{Three spatial generalization test cases.} \textit{Rotate Longitude} rotates all spatial input fields
    (variables and boundary conditions) by 180\textdegree{} in longitude relative to Baseline. \textit{Reverse Latitude} reverses them in latitude; \textit{Reverse Longitude} reverses them in longitude. All are illustrated here using a schematic land-sea mask. Shading is superimposed to reflect the sun's relative position that is unchanged with respect to continents, shown here for 21 June 2025, 6 am UTC. Vector fields are also subject to sign changes, illustrated with a schematic low-pressure system (orange L) originally on the Northern Hemisphere (Baseline).
    For \textit{Rotate Longitude}, all scalar and vector fields are subject to rotation, with velocity components $u$, $v$, and the planetary rotation rate $\Omega$ retaining their sign.
    To retain geostrophic balance, \textit{Reverse Latitude} requires a sign change in the meridional velocity $v$ as the low-pressure system is mirrored to the Southern Hemisphere. 
    The direction of the Coriolis force $\mathbf{f}$ is denoted, the pressure gradient force points inward in all cases. \textit{Reverse Longitude} requires a sign change in the zonal velocity $u$ and in the planetary rotation rate $\Omega$ as all fields are mirrored along the prime meridian 0\textdegree{}E such that the planet effectively rotates in the opposite direction. See Appendix~\ref{proof} for a derivation of the changes across test cases.}
\label{fig:planets}
\end{figure}

\begin{figure}[H]
\noindent\includegraphics[width=\textwidth]{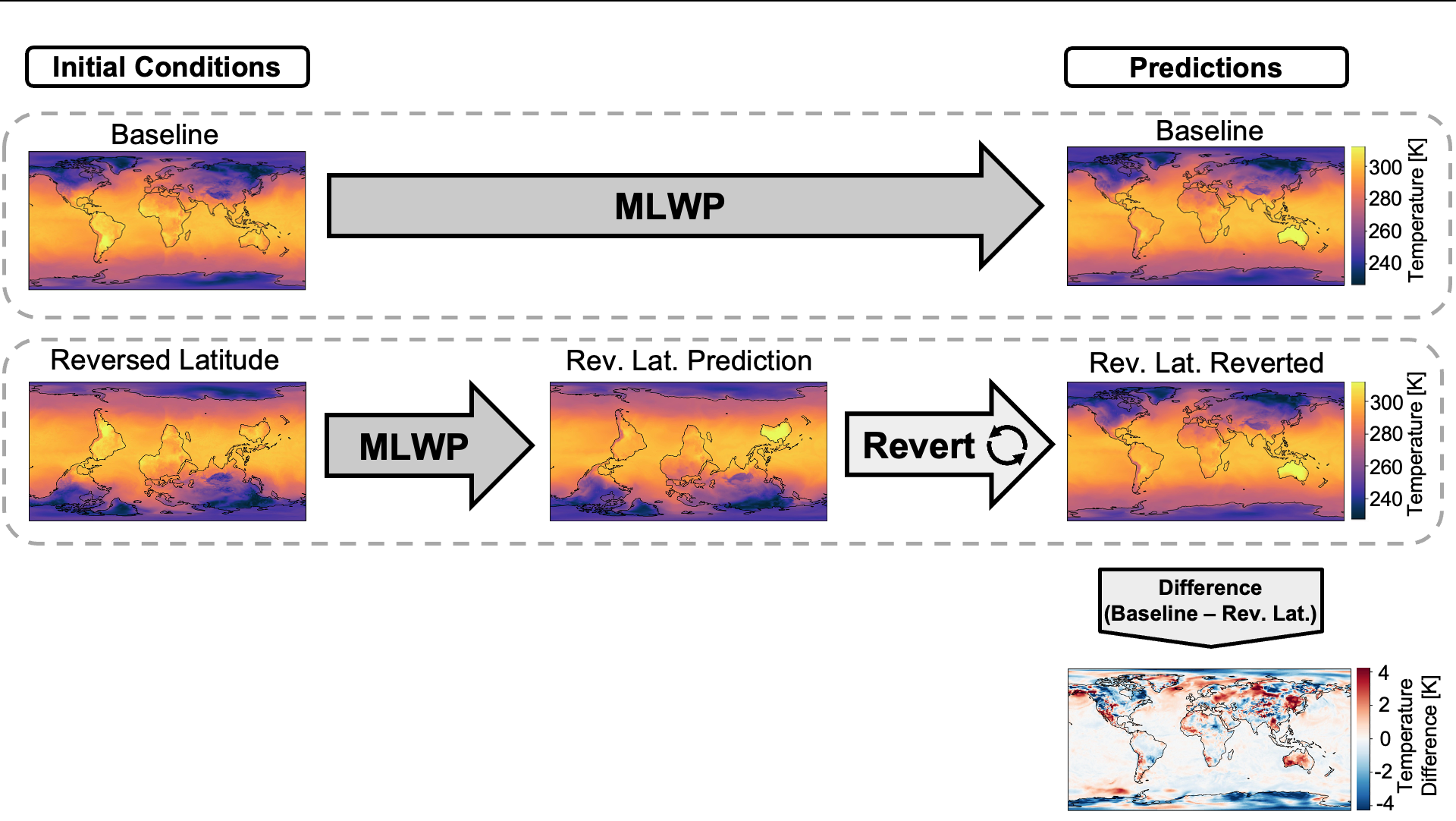}
\caption{\textbf{
    Illustration of the generalization error calculation.} 
    2m-temperature is plotted for illustration but the same process is performed for all other initial and boundary conditions.
    The initial conditions (left column) are used to initialize the machine learning-based weather prediction model (MLWP).
    Advancing the model in time is illustrated using a gray arrow labeled “MLWP".
    A snapshot of the resulting predictions is shown in the right column.
    The first row illustrates the baseline simulation, where all initial conditions are kept in their original configuration.
    The second row illustrates the process for the reversed latitude test case as one example of the three proposed cases.
    When a test case is performed, the transformation that was applied to the initial conditions is reversed in the resulting predictions, as illustrated by the “Revert" arrow to compare errors.
    The bottom right shows the difference between the baseline and reverted reversed latitude predictions, defined as the \textit{generalization error}.}
    \label{fig:process}
\end{figure}

\section{Formulating spatial generalization test cases}

Many MLWP models use coordinate information, such as latitude and longitude as inputs during training and inference.
Such models can learn local climate information based on where on the planet a grid cell is.
This information might increase skill for weather forecasting, which is a prediction problem where stable distributions are a good approximation.
A model can, for example, learn that specific regions are semi-arid in the training data and therefore predict a strong daily cycle (low soil heat capacity due to reduced soil moisture).
However, land-use change or altered precipitation patterns under global warming may modify these regions, resulting in inaccurate predictions when the model overfits on present-day climate.
Learning from coordinates also opens the possibility for non-physical mappings from inputs to predictions, such as learning a daily cycle directly from date-time information rather than from the solar zenith angle, clouds, surface heat fluxes, and local heat capacities.
Such shortcuts may hinder generalization in time and space, as many boundary conditions change naturally or through anthropogenic forcing.

In conventional physics-based general circulation models, this issue generally does not arise.
These models use incoming solar radiation to simulate the daily and annual cycle and are not tuned to specific geographic locations.
Parameters describing vegetation and soil type at a given location are explicitly prescribed in the inputs rather than implicitly inferred from coordinates, as can be the case when MLWP models are trained with coordinate information but without the corresponding physically meaningful surface quantities as predictors.

To investigate the degree to which MLWP models demonstrate spatial invariance, we propose three test cases (Fig.~\ref{fig:planets}), which we perform as illustrated in Fig.~\ref{fig:process}.
These tests imagine forecasting the weather on a planet Earth in one of three different configurations: \emph{Rotate Longitude}, \emph{Reverse Latitude} and \emph{Reverse Longitude}. In these setups, the coordinate system of the Earth is different. For example, in Rotate Longitude the prime meridian (0\textdegree{}E) runs through the Central Pacific instead of Greenwich, London, which was a historic choice but physically arbitrary.
The underlying physical laws remain unchanged in all cases. 

That coordinate transformations do not alter the underlying atmospheric physics can be demonstrated using a physics-based general circulation model.
These models solve the Navier–Stokes equations on a rotating sphere.
The equations are spatially invariant except for advection, Coriolis, and pressure gradient terms, which introduce sign changes (see Appendix~\ref{proof}) that can be formulated simply as sign changes to the initial conditions and the Coriolis parameter.
Thus, when performing coordinate transformations in a general circulation model, appropriate sign adjustments must be applied to yield the same predictions as in the baseline configuration using standard longitude and latitude.
Incoming solar radiation must also be adapted, since the zenith angle calculation (a boundary condition) does depend explicitly on space and time.
Most models calculate the zenith angle at each timestep, but being a boundary condition, we reverse/rotate the resulting field consistently (Fig.~\ref{fig:inis}; shading in Fig.~\ref{fig:planets}).
We demonstrate the spatial invariance of a general circulation model experimentally using SpeedyWeather~\citep{klowerSpeedyWeatherjlReinventingAtmospheric2024} (Fig.~\ref{fig:gen_error}) and derive the spatial invariance of the Navier-Stokes equations (Appendix~\ref{proof}).

The initial conditions (temperature, humidity, winds etc.) and boundary conditions (sea surface temperature, surface geopotential, etc.) need to be transformed into the \emph{Rotate Longitude}, \emph{Reverse Latitude} and \emph{Reverse Longitude} configurations.
To realize these counterfactual Earths in an MLWP model, we rotate all input variables and boundary conditions by 180\textdegree{} in longitude, reverse them in latitude, or reverse them in longitude, while maintaining physical consistency with the aforementioned sign changes (Fig.~\ref{fig:planets}).
Here, physical consistency means that the transformed configuration should yield the same predictions as the baseline case, once its transformation is reverted.
For example, if the initial conditions are rotated by 180\textdegree{} in longitude, the resulting forecast corresponds to this counterfactual configuration;
rotating the forecast back by 180\textdegree{} should recover the baseline prediction up to rounding errors, if the model generalizes spatially and all necessary adaptations for physical consistency have been made.

As time is used as a predictor in some MLWP models (e.g., GraphCast or GenCast) changes apply here too for consistency:
When rotating in longitude by 180\textdegree{}, incoming solar radiation corresponds to a 12-hour shift in the daily cycle.
Reversing latitude shifts seasonality by half a year (shading in Fig.~\ref{fig:planets}, \emph{Reverse Latitude}) in addition to a sign change for meridional velocity, $v \to -v$.
For the reversal in longitude, additionally, the sign of the Coriolis force changes because planetary rotation must be reversed to retain geostrophic balance when the zonal velocity is negated, $u \to -u$ (Appendix~\ref{proof}).

We evaluate each MLWP model’s forecast skill under these three transformations.
A deterministic model that generalizes spatially should produce predictions identical to the baseline case, only subject to a rounding error, after reversing the transformation.
A model is considered to generalize spatially if its generalization error is much smaller than its baseline forecast error.
Here, the generalization error for a model is comparing the baseline simulation with the transformed one (Fig.~\ref{fig:process}), whereas the forecast error is comparing the baseline simulation to reanalysis data used as reference.

These tests pose a straightforward way to investigate the generalization of a model's simulated physics in space.
There are many possible causes for a lack of spatial generalization, including the model architecture not being invariant or utilization of predictors that include coordinate information.
Conducting these tests can flag any unintended spatial dependencies, potentially disentangle variable mappings, and --- in the future --- could be used as a training signal for developers to increase spatial generalization in their MLWP models.
Overall, we acknowledge that providing coordinate information can be viewed as a combined “parameterization" for several sub-grid scale processes and implicitly represented boundary conditions.
However, making such a parameterization location-dependent may hinder generalization, and cause physically inconsistent mappings from inputs to outputs.
This inconsistency is detrimental when trusting these models with the climate prediction task.

\section{Results}
\subsection{Error growth}\label{sec:error_growth}
\begin{figure}[H]
\noindent\includegraphics[width=\textwidth]{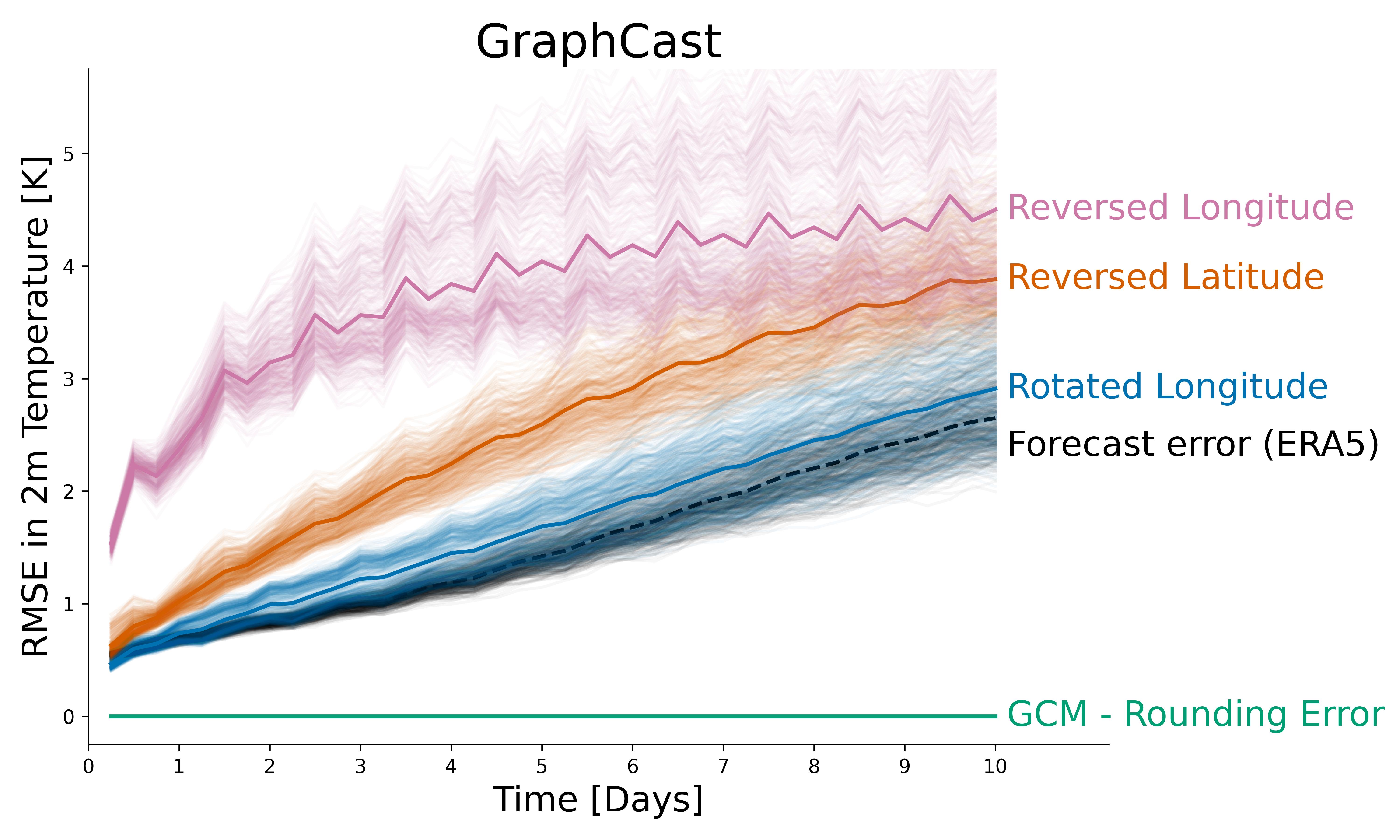}
    \caption{\textbf{Spatially averaged generalization errors in 2m-temperature of GraphCast small.} Generalization errors from the three test cases are indicated in pink, orange, and blue.
    Thin lines are ensemble member errors, thick lines the ensemble mean errors. The black lines indicate the forecast error of the same model in Baseline configuration (no reverse/rotate) compared against ERA5 reanalysis data.
    The ensemble is obtained from 365 simulations, initialized at midnight on each day of the year 2022.
    Green lines indicate the generalization errors for all three test cases in the physics-based general circulation model (GCM) SpeedyWeather, only resulting from rounding errors
    (see Fig.~\ref{fig:SpeedyErrors} for more details).
    All errors are area-weighted root-mean square errors (RMSE, see Section~\ref{error_calc}).
    }
    \label{fig:gen_error}
\end{figure}

We apply our test cases to the GraphCast small MLWP model, which has a 1\textdegree{} resolution and is the least computationally intensive of the available GraphCast versions~\citep{lamLearningSkillfulMediumrange2023}.
Otherwise the model architecture is identical to GraphCast versions at other resolutions.
GraphCast uses latitude and longitude coordinates, as well as date-time information, as inputs to make predictions, which makes it easier for the model to overfit on coordinate information.
The average \textit{generalization error} (see Fig.~\ref{fig:process}) in 2m-temperature of GraphCast is larger than the \textit{forecast error} of the model
(Fig.~\ref{fig:gen_error}, for details on the error calculation see Section~\ref{error_calc}).
Hence, errors in generalization are not negligible compared to the forecast errors and we therefore conclude that GraphCast does not generalize spatially.
In contrast, the test cases in the physics-based general circulation model SpeedyWeather~\citep{klowerSpeedyWeatherjlReinventingAtmospheric2024} only result in a rounding error (Fig.~\ref{fig:SpeedyErrors}), which at sufficient precision is negligible compared to any forecast error~\citep{dubenBenchmarkTestsNumerical2014, chantryScaleSelectivePrecisionWeather2019}.

GraphCast fails each generalization test, but by a different margin.
The rotated longitude test case produces, on average, only a slightly larger generalization than forecast error,
whereas the reversed longitude case has, depending on the forecast lead time, on average, a 1.5-3 times larger generalization error.
This difference in error magnitude can already be seen after the first 6-hour time step.
For Rotate Longitude, the generalization error after 6 hours is smaller than the forecast error at this lead time, but grows faster.
The error for reversed longitude case, in contrast, is already above 1K after 6 hours, or approximately equal to the forecast error after 5--6 days.
Overall, we observe different levels of difficulty for each test case, presumably related to their difference from Baseline.
In terms of changes to the simulated physics, Rotate Longitude would be the easiest test case to pass, followed by Reversed Latitude then Reversed Longitude.
Whether these difficulty levels generally apply across a range of MLWP models is unclear.

\begin{figure}[H]
\noindent\includegraphics[width=\textwidth]{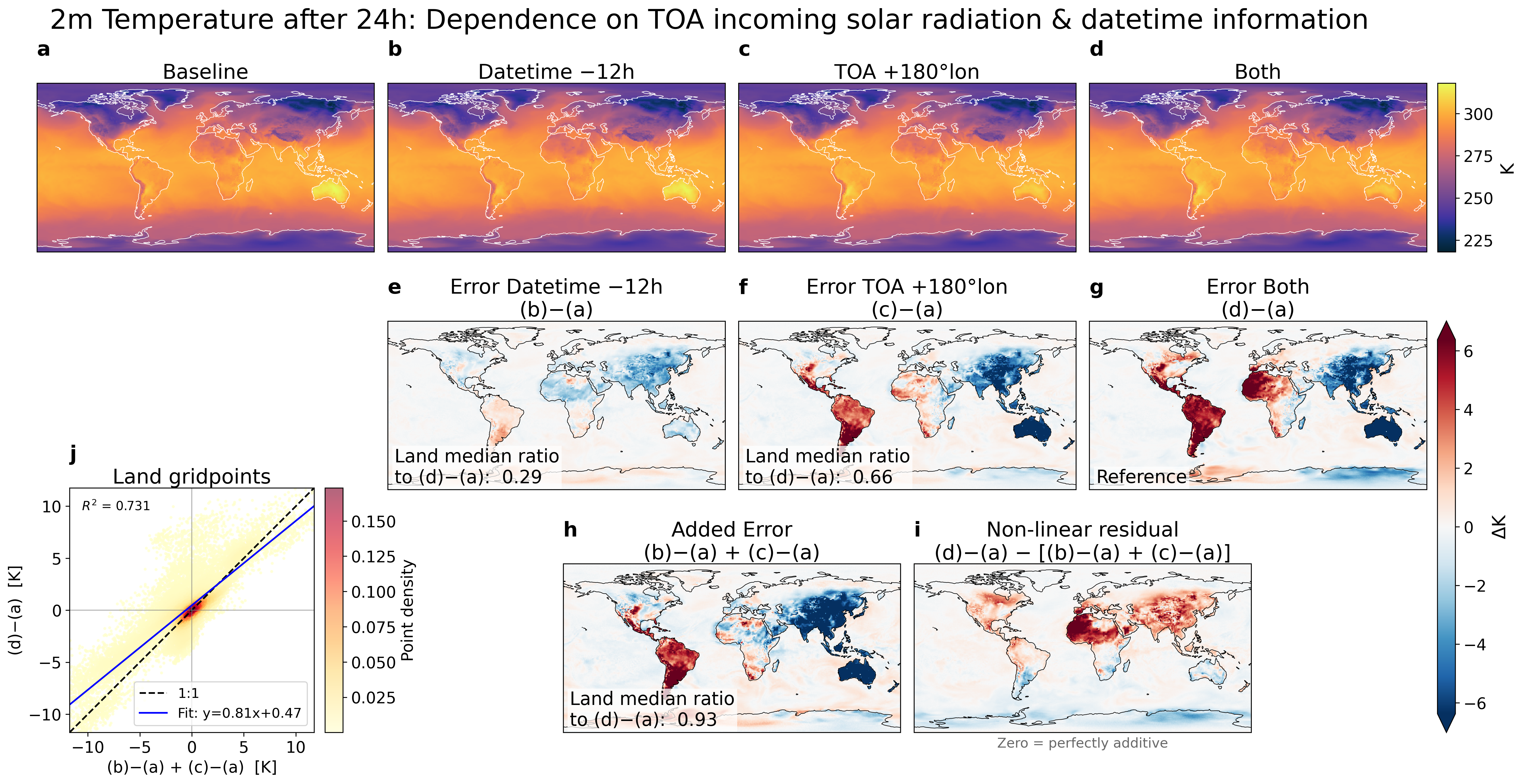}
    \caption{\textbf{Daily cycle learned from combined solar radiation, time and coordinate information}. 
    (a) GraphCast 24-hour forecast of 2m-temperature initialized on 2022-01-01,
    (b) as (a) but 12 hours were subtracted from date-time input to GraphCast,
    (c) as (a) but incoming solar radiation shifted by 180\textdegree{} in longitude,
    (d) as (a) but both changes from (b) and (c) are applied.
    (e-g) are the respective errors of (b-d) quantified as difference to (a).
    (h) For a linear model, this error decomposition would add up (h identical to g),
    but (i) the non-linear residuals of this decomposition are non-zero.
    (j) The correlation of the linear error decomposition.
    To quantify the average contribution of (e, f) to the total error (g) the ratio over land is calculated.
    The median error ratio is 0.29 over land for (e, subtracting 12 hours) and 0.66 for (f, rotating solar radiation by 180\textdegree{}) compared to the error for both changes (g).
    }
\label{fig:dependence_on_TOA_and_datetime}
\end{figure}

A slight daily cycle can be observed in all GraphCast generalization and forecast errors, but it is considerably more pronounced in the reversed longitude case (Fig.~\ref{fig:gen_error}).
GraphCast has both date-time and top-of-the-atmosphere incident solar radiation (i.e., solar constant times cosine of solar zenith angle) variables available to predict the daily cycle.
By initially changing only one of these predictors in the rotated longitude test case, we reveal that GraphCast has learned the daily cycle from a combination of both predictors and coordinate information (Fig. ~\ref{fig:dependence_on_TOA_and_datetime}, ~\ref{fig:dependence_on_TOA_and_datetime_average}).

The contribution of the incident solar radiation to the learned daily cycle is about 2/3, whereas the time information contributes about 1/3, quantified as the median error over land (Fig. ~\ref{fig:dependence_on_TOA_and_datetime}e, f).
Arguably, the time-of-day should not influence temperature at all, given that it is only correlated with, but not the cause of, daily temperature variations.
As such, learning the daily cycle partly from combined date-time and coordinate information, rather than exclusively from incident solar radiation, provides a non-physical mapping from inputs to outputs.

\subsection{Spatial distributions}

\begin{figure}[H]
\noindent\includegraphics[width=\textwidth]{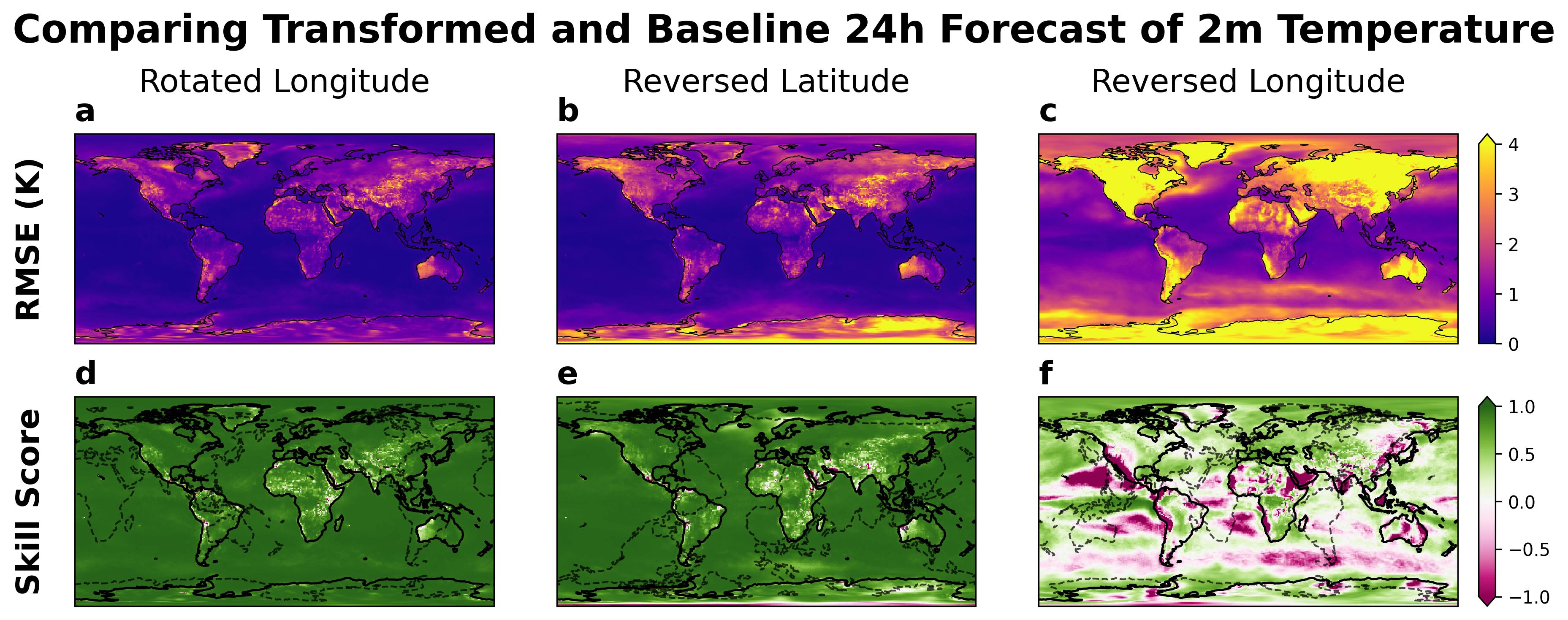}
    \caption{
    \textbf{Spatial distributions of errors in 2m-temperature forecasts of GraphCast small over 24 hours.}
    (a) RMSE of Rotated Longitude test case compared to Baseline, 
    (b) as (a) but for Reversed Latitude,
    (c) as (a) but for Reversed Longitude.
    (d-f) as (a-c) but for the climatological skill score.
    Each error map is averaged over 365 simulations with a 24-hour lead time, initialized at midnight on each day of 2022.
    The climatological skill score (higher is better) uses ERA5 reanalysis (1992-2022) for the reference mean squared errors (Equation~\ref{eq:SS} in Section~\ref{error_calc}).
    Solid coastlines indicate Baseline, i.e. after the test case transformations have been reversed.
    Dotted lines indicate the transformed coastlines.
    }
    \label{fig:spatial_distr}
\end{figure}

For the rotated longitude and reversed latitude test cases, the generalization error is largest in mountainous regions, as shown by the climatological skill score (Fig.~\ref{fig:spatial_distr}).
In these locations, GraphCast relies on localized coordinate information and not entirely on physical boundary conditions such as the surface geopotential (orography times gravity),
which is an available predictor for the model.
GraphCast has correctly learned that atmospheric temperatures generally decrease with height, so changes to the surface geopotential immediately imprint on 2m-temperature (Fig.~\ref{fig:dependence_on_geopot_and_lon}).
However, one reason for GraphCast to rely on coordinate information regardless could be insufficient predictors for other processes in these regions.
Physics-based models, for instance, utilize additional inputs such as the standard deviation and slope of the sub-gridscale orography to better represent local boundary layer processes.
By including these boundary conditions explicitly, a model might be able to learn physically consistent mappings from input to output data, without relying on coordinate information, which inevitably limits the ability to adapt to changing boundary conditions.

At the same time, GraphCast seems to have learned a physically consistent mapping from the land-sea-mask to 2m-temperature.
In the rotated longitude case, for instance, there are only small errors over Africa despite the continent essentially being moved to the coordinates of the Pacific Ocean (Fig.~\ref{fig:spatial_distr}a and d).
In the reversed longitude case, errors are considerably larger than in the other two test cases. 
As GraphCast has learned the daily cycle partly from date-time information (Fig.~\ref{fig:dependence_on_TOA_and_datetime}) we adjust the other test case by shifting the time by half a day or year.
However, in the reverse longitude test case, this time adjustment is not possible as time would either need to run backwards or be formulated as a local, coordinate-dependent time for consistency with the daily cycle. 
Neither interventions are supported by the GraphCast inputs, so time is left unchanged and therefore inconsistent with other predictors in Reverse Longitude. 
We do not consider this a shortcoming of our test cases but a consequence of the unphysical reliance of the model on coordinate and time information.
Additionally, the Coriolis parameter is learned rather than prescribed in GraphCast, causing an inevitable physical inconsistency in the Reversed Longitude test case.

\subsection{NeuralGCM}
\begin{figure}[htbp]
    \centering        \centering\includegraphics[width=0.95\textwidth]{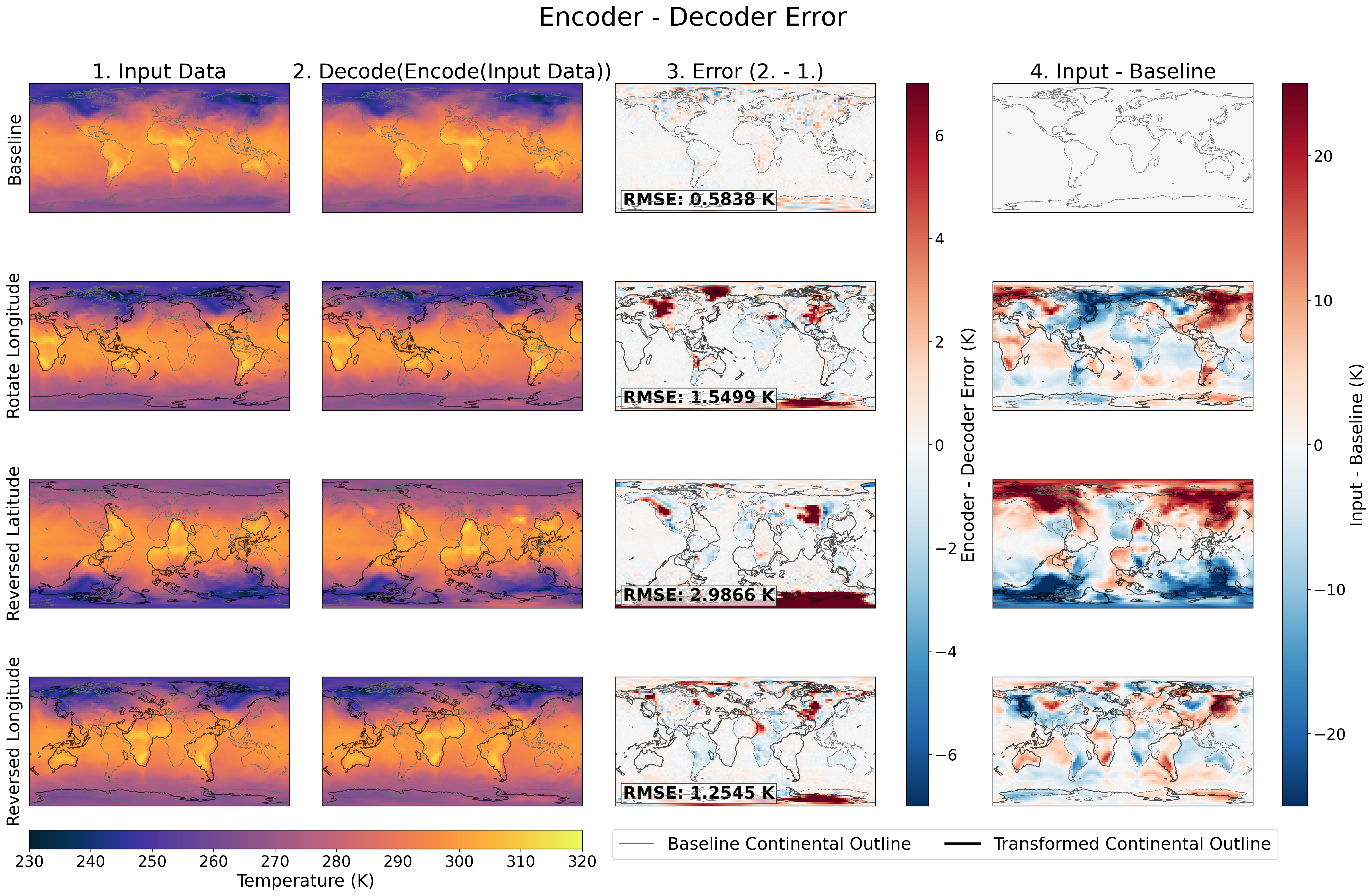}
        \caption{
        \textbf{Spatial generalization of NeuralGCM's encoder-decoder.}
        ERA5 temperature at the lowest level is encoded and decoded for each test case (rows), no time integration is applied.
        Column 1: Temperature inputs in each test case.
        Column 2: Temperature after encoding and decoding.
        Column 3: Encoder-decoder error as the difference (Column 2 minus 1). The area-weighted average RMSE is denoted.
        Column 4: Difference between the temperature inputs for the respective test case and the baseline temperature input.
        Black coastlines are for the transformed input data, gray coastlines are untransformed.}
        \label{fig:encode_decode}
\end{figure}
NeuralGCM is a hybrid general circulation model, with a physics-based dynamical core and machine-learned parameterizations.
The model uses vertical sigma coordinates (fraction of surface pressure) internally~\citep{kochkovNeuralGeneralCirculation2024}.
To convert the ERA5 data, which uses pressure coordinates, to sigma coordinates, as well as to apply a learned correction (used to filter out gravity waves from initialization shocks) NeuralGCM uses an encoder - decoder module in its architecture~\citep{kochkovNeuralGeneralCirculation2024}.
Among other variables, this learned module uses the sine and cosine of latitude and longitude as well as additional orography features and positional embeddings as inputs.
When applying the generalization test to this model, we found that simply encoding and then decoding the input data, without any time integration, already does not generalize spatially. 
Additional continental features appear in the lowest-level temperature where continents were before the coordinate rotation/reversal was applied, termed “ghost continents" (Fig.~\ref{fig:encode_decode}).
Across test cases, the errors after encoding and decoding temperature typically double in each test case compared to the error without coordinate rotation/reversal.
Many error features take the form of ghost continents and therefore indicate reliance on coordinate information as predictors, instead of adapting to new initial conditions in a physically consistent way.
There is not just a higher encoder-decoder error on land, but specifically at the locations where the NeuralGCM encoder-decoder \emph{expects land to be}.
As with GraphCast, the errors are frequently located in mountainous regions.
The largest errors are in the reversed latitude case with the reversal of the Arctic and Antarctic regions, coinciding with an altitude change of 2000-3000m in many areas.
Notably, here, a reversal in latitude results in higher errors than a reversal in longitude, as there is a larger area mismatch in local temperatures between the baseline and the test case for the latitude reversal.
While we focus here on the encoder-decoder module of NeuralGCM, including a time integration is unnecessary as errors will only be further amplified.
Spatial generalization needs to hold for every model component for physical consistency. 
To evaluate the spatial generalization of the forecasting component of the model, one could perform a manual conversion between pressure and sigma coordinates, although this would exclude the learned correction of NeuralGCM.

\section{Discussion}

We investigate whether machine-learned weather prediction models behave unphysically when extrapolated to our proposed spatial generalization test cases.
These limitations can for instance arise from hardcoded representations of spatial and temporal relationships in the training data in their latent space,
or unphysical predictors like time or coordinate information for processes that are space and time invariant.
We show that currently, the fully data-driven model GraphCast, and the hybrid general circulation model, NeuralGCM, do not pass these test cases.
For GraphCast we show that it has learned unphysical relationships based on coordinate and time information, instead of the more general underlying atmospheric physics principles.
For NeuralGCM's encoder-decoder we observe so-called \emph{ghost continents}, i.e., features aligned with where the model has learned continents to be, rather than solely inferred from orography and the land-sea mask.
These architectural choices pose a challenge to spatial generalization, questioning the ability of such models to simulate a changing regional climate. 

While the predictive power of a weather forecast model is its most important feature, other aspects such as trustworthiness, interpretability, and generalizability are required for models to be applied for climate predictions.
Model architectures that respect spatial invariance can be advantageous even at shorter weather time-scales.
Lam et al.~\cite{lamLearningSkillfulMediumrange2023}, for instance, state that GraphCast benefits from retraining with more recent data.
A model that generalizes spatially, should be more robust to changing climatic conditions and hence might also require less frequent retraining, making it an overall more robust model.
While it may appear advantageous to use coordinate information as a “parameterization" for local climate processes, e.g. to learn land-surface types or sub-grid-scale orography,
this prevents using the model for much of scientific experimentation.
For instance, the effects of land-use change, melting sea-ice, or idealized solar forcing experiments, can not be conducted with such a model.

In GraphCast, we identified particularly high errors in mountain ranges and linked this to over-reliance on coordinate information instead of solely orography. 
Similarly, GraphCast has learned the daily cycle based on the date-time information, rather than exclusively from the cosine of the zenith angle.
Both relationships are considered unphysical \emph{shortcuts} that violate generalization.
While the length of day only changes on paleo time-scales, this points to an underlying problem of MLWP architectures whereby predictors can be found for which physically there should not be any dependence.

Our proposed test cases were able to identify different shortcomings in two machine-learned weather models with regard to spatial generalization.
In the future, research should be conducted to verify and empirically support the claim of spatial invariance as a necessary but not sufficient condition for temporal invariance.
Furthermore, different model architectures can be evaluated for spatial invariance, as some explicitly enforce symmetries or do not use coordinate information in the training data.
Additionally, training on spatially transformed training data could yield improvements in this regard.
Overall, we recommend the adoption of the presented spatial invariance tests during model development to ensure that next-generation weather and climate models are not only accurate but also physically consistent and robust under climate change scenarios.

\section{Methods}

\subsection{Test Execution}\label{execution_in_practice}
We propose to conduct our tests on an MLWP model following the practical guidelines in Table~\ref{tab:test_instructions}. We ourselves apply them to the MLWP models NeuralGCM and GraphCast, as well as the atmospheric general circulation model SpeedyWeather as described in Table~\ref{tab:practice_NGCM_GC}. 
\newpage
\begin{center}
\begin{tblr}{
  colspec = {X[t,c,3.5cm] X[c,h,8cm]},
  stretch = 0,
  rowsep = 6pt,
  hlines = {black,1pt},
  vlines = {black,1pt},
}
Rotation or reversal & Applied to \\
Mandatory & 
\begin{itemize}[topsep=0pt, partopsep=0pt, parsep=0pt, itemsep=0pt, left=0pt]
    \item All initial and boundary conditions
\end{itemize} \\
If possible &
\begin{itemize}[nosep, left=0pt]
    \item Solar zenith angle calculation or field
\end{itemize} \\
Disallowed &
\begin{itemize}[nosep, left=0pt]
    \item Coordinate information (unless done instead of rotating initial \& boundary conditions)
    \item Embeddings
\end{itemize} \\
\end{tblr}
\captionof{table}{Guidelines for performing the spatial generalization tests on a general MLWP model. 
Additionally, some variables need to be adapted, if possible (depending on model architecture, e.g. models may not provide an interface to change Coriolis), and some should not be changed. 
Generally, the model's internal coordinates and structure (including embeddings) should remain unchanged but inputs should be adapted according to the test cases. 
If the input to the model is an xarray, rotating or reversing the coordinate information may be equivalent to reversing or rotating each of the boundary and initial conditions.
Some wind components (u, v, or vorticity) and the Coriolis parameter also need to be negated, depending on the test case, as shown in Fig.~\ref{fig:planets}.
}
\label{tab:test_instructions}
\end{center}

All initial and boundary condition fields are subject to rotation or reversal.
Additionally, for external forcings like the solar zenith angle which is in many models computed at each timestep, either the calculation itself or the resulting field should be adapted if possible.
The planetary rotation needs to be negated for Reversed Longitude, if models allow this.
Additionally, wind vectors need to be changed as illustrated in Fig~\ref{fig:planets}.
Generally coordinate information should remain unchanged such that the model's internal operators act on rotated/reversed fields. 
However, depending on the data structure of the model input, for instance for xarrays, rotating or reversing the coordinate information may be equivalent to reversing or rotating each of the boundary and initial conditions.
Adapting all of a model's embeddings defeats the point of testing spatial generalization of the simulated physics, and instead makes it a test of the model architecture beyond embeddings. After all, the tests aim to highlight the shortcomings of hardcoding spatial information in the first place.

\begin{center}
\begin{tblr}{
  colspec = {X[t,c,4cm] X[c,h,8cm]},
  stretch = 0,
  rowsep = 6pt,
  hlines = {black,1pt},
  vlines = {black,1pt},
}

SpeedyWeather, NeuralGCM &
\begin{itemize}[nosep, left=0pt]
    \item Initial and boundary conditions
    \item Solar zenith angle calculation
    \item Planetary rotation
    \item Wind components
\end{itemize} \\
GraphCast &
\begin{itemize}[nosep, left=0pt]
    \item Initial and boundary conditions
    \item Including incident solar radiation
    \item Input time subtracted 12 hours (Rotate Longitude), added 183 days (Reverse Latitude)
    \item Wind components
\end{itemize}  \\
\end{tblr}
\captionof{table}{Variables adapted for each model in practice.
For SpeedyWeather and NeuralGCM (encoder-decoder only) we adapt the models as outlined in Table~\ref{tab:test_instructions}.
For GraphCast, planetary rotation or Coriolis force is learned and not changeable.
Additionally, time is a model input and therefore adapted to be consistent with the changes to the incident solar radiation.
}
\label{tab:practice_NGCM_GC}
\end{center}

\begin{figure}[htbp]
    \noindent\includegraphics[width=\textwidth]{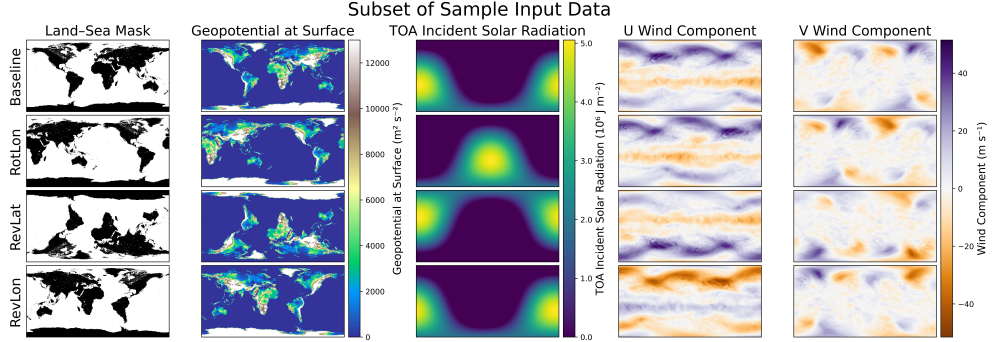}
    \caption{Example of a subset of GraphCast initial conditions for the three test cases, including sign changes for the wind components.
    In ERA5, top-of-atmosphere (TOA) incident solar radiation is accumulated, units here are integrated Wm$^{-2}$ over one hour.
    Surface geopotential is orographic height times gravity.}
    \label{fig:inis}
\end{figure}

In practice, these guidelines can be ambiguous, depending on the model (Table ~\ref{tab:practice_NGCM_GC}).
For instance, while the incident solar radiation is a boundary condition it is typically computed during a model integration.
But as a boundary condition, it is subject to rotation/reversal.
GraphCast, in contrast, also takes time as an additional input.
While we argue that an explicit dependency on time is not physical, we adapt the input time to stay consistent with changes made to the top-of-atmosphere incident solar radiation.
For Rotate Longitude 12 hours have been subtracted to match the time of the daily cycle; 183 days are added for Reverse Latitude to match the seasonal cycle.
To maintain physical consistency in the Reverse Longitude simulations, time would either need to run backwards or be formulated as a local time depending on spatial coordinates.
Both are not possible with GraphCast's user interface so we leave time unchanged in this test case.
This leads to higher expected errors for Reverse Longitude but is considered a shortcoming of GraphCast's dependency on time, not of the test case itself.
Wind components are changed in all models ($v \to -v$ in Reverse Latitude, $u \to -u$ and planetary rotation $\Omega \to -\Omega$ in Reverse Longitude).
Relative vorticity is negated instead of $u, v$ in SpeedyWeather, which uses a vorticity and divergence as variables for the horizontal wind, see Appendix~\ref{proof}.
Rotation and reversal of some GraphCast's initial conditions is illustrated in Fig.~\ref{fig:inis}.

\subsection{Mathematics of rotation and reversal}\label{sec:math_background}
Rotate Longitude, Reverse Latitude, and Reverse Longitude mathematically correspond to reflection
(with respect to the Equator 0\textdegree{}N or Prime Meridian 0\textdegree{}E) and half-turn rotation (by 180\textdegree{}E).
As such, these cases are examples of mathematical involutions, functions that are their own inverse.
Applying any of these twice yields Normal (the baseline configuration) again.
Call the three test cases $A, B, C$ for brevity, and Normal $\mathbbm{1}$, then $AA = BB = CC = \mathbbm{1}$ and note that while we focus on $A, B, C$ for simplicity,
also four more test cases $AB, AC, BC, ABC$ can be constructed. Note that given commutativity
we have $AB=BA$ etc.
In essence, while we propose three basic test cases for spatial generalization there are
seven in total. But many more can be constructed using, for example, quarter turns by 90\textdegree{}E or other degrees as necessary on some grids.

\subsection{Rotation and reversal on discrete grids}\label{sec:grid_explainations}
Variables defined on many grids, i.e. discretizations of the sphere, can be rotated or reversed by reordering the elements in the underlying arrays intuitively. 
For a grid with longitudinal points starting at 0\textdegree{}E (no offset), the grid point order for Reverse Longitude might be counterintuitive as the grid point at 0\textdegree{}E is excluded from the re-ordering.
For example, $a_1, a_2, a_3, a_4$ (at 0\textdegree{}, 90 \textdegree{}, 180\textdegree{}, 270\textdegree{}) would be reversed to $a_1, a_4, a_3, a_2$ to avoid an additional rotation by the grid spacing.
For grids that are symmetric around the equator and the prime meridian, the reversals in the respective dimensions are exact. Without symmetry, interpolation is necessary.
Similarly, to rotate the longitude by 180\textdegree{}E, a rotational symmetry of an even order is required.
Grids based on an icosahedron (e.g. ICON and GraphCast) have a rotational symmetry of order 5, such that variables on those grids can only be rotated exactly by multiples of 72\textdegree{}.
We suggest here to adapt the Rotate Longitude test case accordingly, e.g. by rotation of 144\textdegree{} in longitude instead.
Note that GraphCast employs a learned encoder-decoder architecture such that the input data is not provided on an icosahedral but a regular longitude-latitude grid.
The cubed sphere (e.g. MITgcm) and the HEALPix grid~
\citep{gorskiHEALPixFrameworkHighResolution2005} have a rotational symmetry of order 4, allowing both to rotate longitude by 180\textdegree{} without interpolation.
For regular longitude-latitude grids the number of grid points in longitude has to be divisible by 2 but otherwise rotation by different amount of degrees is possible.

\subsection{Rotation and reversal in spherical harmonics}\label{sec:grid_explainations_spherical}

Many physics-based atmospheric general circulation models, as well as some ML-based ones, use spherical harmonics to represent variables spatially.
Rotating a variable $a_{l,m}$, represented as coefficient of the spherical harmonic of degree $l$ and order $m$, in longitude by $\lambda$ (angle in radians) is $\exp(-im \lambda)a_{l,m}$, which is a rotation in the complex plane of every coefficient respectively.
Reversing latitude is a negation of the odd harmonics, i.e. $-a_{l,m}$ if $l+m$ odd.
Reversing longitude is the complex conjugate of every coefficient $\bar{a}_{l,m}$.

\subsection{Generalization error calculation}\label{error_calc}

First, we ran 10-day forecasts initialized from midnight at each day of the year 2022 (and the last day of the year 2021) in the regular GraphCast small set-up.
GraphCast small is the least computationally intensive of the available GraphCast versions at a 1\textdegree{} instead of 1/4\textdegree{} resolution and with 13 pressure levels~\citep{lamLearningSkillfulMediumrange2023}.
From these runs, we were able to calculate the GraphCast \textit{forecast error} at each timestep.
This is done by calculating the RMSE in 2m-temperature between the GraphCast predictions in the baseline setup and the ERA5 data for the same day.
We chose 2m-temperature, since this variable was weighted more highly in GraphCast's loss function than other surface variables~\citep{lamLearningSkillfulMediumrange2023}.
Furthermore, 2m-temperature has a strong signal in the daily and seasonal cycles, important for parts of our analysis.

Next, we ran our test cases for the same start dates and calculate the \textit{generalization errors}.
These are the RMSEs in 2m-temperature between GraphCast's baseline with the rotated/reversed test case (see Fig.~\ref{fig:process}).
The generalization error is therefore computed from a model with itself (in rotated/reversed form) and not in comparison against ERA5.
We are reversing the respective test case transformation after the predictions are made to compare corresponding regions (Fig.~\ref{fig:process}).
The generalization error is calculated at each timestep and the area-weighted spatially averaged error is analyzed (Fig.~\ref{fig:gen_error}).

For the climatological skill score, we calculate the MSE between the baseline and adapted setups, divide it by the MSE between the 30-year climatological mean at each location at each 6h timestep throughout the year in ERA5, with the corresponding ERA5 values at each 6-hour time step in 2022, and subtract the result from one (Equation~\ref{eq:SS}).  
\begin{equation}\label{eq:SS}
    Skill Score(\tau) = 1 - \frac{\text{MSE}_{\text{baseline, adapted}}(\tau)}
    {\text{MSE}_{\text{ERA5 30yr\_clim, ERA5 2022}}(\tau)}
\end{equation}
Where $\tau$ is the lead time. The results for $\tau$ = 24h are shown in Fig.~\ref{fig:spatial_distr}.

For our test, we stipulate that if the \textit{generalization error} is much smaller than the \textit{forecast error} of the model, the model generalizes spatially.
\begin{equation}\label{eq:generalization_error}
\text{Error}_{Generalization} << \text{Error}_{Forecast}
\end{equation}

We do recognize the limitations of this criterion, in that a model with a low forecast error will have a harder-to-reach threshold for spatial generalization.
However, the generalization error can be on the order of the rounding error and therefore very small which means that the exact formulation is less important
than conceptually that a generalization error should be negligible compared to the forecast error.
We expect that for most models the spatio-temporal analysis of the generalization error will be the most insightful,
rather than a binary classification into pass or fail alone.

\appendix

\section{Spatial invariance of the Navier-Stokes equations}\label{proof}

An atmospheric model solving the discretised Navier-Stokes equations passes the spatial generalization tests in this study because the Navier-Stokes equations are invariant under reversal and rotation. We demonstrate this here.

\subsection{Reverse Latitude or Longitude}

Let $R$ be the reversal or rotation operator in general, and be $R_x, R_y, R_o$ in the particular case of Reverse Longitude,
Reverse Latitude, and Rotate Longitude, respectively. We will discuss the Rotate Longitude as a special case later.
As mathematical involution we have $RR = \mathbbm{1}$ (see Section~\ref{sec:math_background}). As an operator applying spatial reordering only,
$R$ has the following linear properties with respect to all element-wise operations on a scalar or vector field $u$, i.e.
$R(u + v) = Ru + Rv$, $R(uv) = R(u)R(v)$ for another field $v$, $R(au) = aRu$ for a spatial constant $a$.
Consequently, we also have $R(uR(v)) = (Ru)v$ and commutativity with the element-wise inverse $R(\tfrac{1}{u}) = \tfrac{1}{Ru}$.
Gradient operators are negated if the reversal is in the same direction $\partial_x R_x u = -R_x\partial_x u$,
$\partial_y R_y u = -R_y\partial_y u$ (anticommutative), but $\partial_x R_y$ and $\partial_y R_x$ are commutative.
Consequently, we have $\partial_x^2 (Ru) = R\partial_x^2 u$ as the second derivatives undo
any sign change from the first derivatives and so $R$ is also commutative with the Laplace operator
$\nabla^2 (Ru) = R\nabla^2 u$.

The two-dimensional incompressible Navier-Stokes equations with Coriolis force can be written as
\begin{equation}
    \partial_t \mathbf{u} + (\mathbf{u} \cdot \nabla) \mathbf{u} + f\mathbf{k}
    \times \mathbf{u} = -\frac{1}{\rho}\nabla p + \nu \nabla^2\mathbf{u} + \mathbf{F}.
    \label{eq:NS}
\end{equation}
$\mathbf{u} = (u, v)$ is velocity, $\mathbf{k}$ is angular velocity, $f\mathbf{k} \times \mathbf{u} = (fv, -fu)$ is Coriolis,
$\rho$ is density, $p$ is pressure, $\nu$ is viscosity, and $\mathbf{F}$ is forcing.
Our goal is to demonstrate the equivalence of (i) solving Eq. \ref{eq:NS} using reversed inputs (initial conditions, parameters and forcing) followed by reversing the solutions back and (ii) solving the original equation with no reversal applied.
The horizontal boundary conditions are assumed to be periodic.
We apply $RR = \mathbbm{1}$ to each term in Eq. \ref{eq:NS}.
We now aim to move one $R$ through the operators to act directly on the inputs, to represent reversing the input and boundary fields.
Simultaneously, we aim to leave one $R$ applied to all terms, to reverse the solution back after the simulation.
We introduce the following notation, $Ru = \overset{\leftrightarrow}u$ in general and $R_xu = \overset{\leftarrow}{u}, R_yu = \overset{\downarrow}{u}, R_ou = \overset{\circlearrowleft}{u}$ in the particular cases.

Starting with the temporal derivative we have
$\partial_t u  = RR \partial_t u = R \partial_t \overset{\leftrightarrow}{u}$,
so using reversed initial conditions $\overset{\leftrightarrow}u$ and then reversing back the solution
are equivalent to solving the unreversed equations because $R$ and $\partial_t$ are commutative. Similar for viscosity, 
$RR \nu \nabla^2 u = R\nu \nabla^2 \overset{\leftrightarrow}{u}$ and forcing
$RR\mathbf{F} = R\overset{\leftrightarrow}{\mathbf{F}}$.
So these terms are spatially invariant under $R$.
For the advection term note the anticommutativity $R_x(u\partial_x u) = -\overset{\leftarrow}{u} \partial_x \overset{\leftarrow}{u}$,
so the sign reverses when the gradient is taken in the same direction as the
reversal operator. Consequently, the advection term becomes
\begin{equation}
R_xR_x (\mathbf{u} \cdot \nabla)\mathbf{u} =
R_x\m{-\overset{\leftarrow}{u}\partial_x \overset{\leftarrow}{u} + \overset{\leftarrow}{v}\partial_y \overset{\leftarrow}{u} \\
      -\overset{\leftarrow}{u}\partial_x \overset{\leftarrow}{v} + \overset{\leftarrow}{v}\partial_y \overset{\leftarrow}{v}},
      \quad R_yR_y (\mathbf{u} \cdot \nabla)\mathbf{u} =
R_y\m{\overset{\downarrow}{u}\partial_x \overset{\downarrow}{u} - \overset{\downarrow}{v}\partial_y \overset{\downarrow}{u} \\
      \overset{\downarrow}{u}\partial_x \overset{\downarrow}{v} - \overset{\downarrow}{v}\partial_y \overset{\downarrow}{v}},
\end{equation}

so the sign changes on the first (second) terms for Reverse Longitude (Latitude) in comparison to the
conventional $(\mathbf{u} \cdot \nabla)\mathbf{u}$ term.
For the Coriolis term $(fv, -fu)$ we note that $R_xf = f, R_yf = -f$ as $f$ is a function of latitude
only and is thus antisymmetric. It follows that $R_xR_x (fv, -fu) = R_x (f \overset{\leftarrow}{v}, -f \overset{\leftarrow}{u})$
so the Coriolis term is invariant under Reverse Longitude but undergoes a sign change for Reverse Latitude,
$R_yR_y (fv, -fu) = R_y (-f \overset{\downarrow}{v}, f \overset{\downarrow}{u})$.
The pressure gradient term becomes
$R_xR_x (-\tfrac{1}{\rho} \nabla p) =
-\tfrac{1}{\overset{\leftarrow}{\rho}}(-\partial_x \overset{\leftarrow}{p}, \partial_y \overset{\leftarrow}{p})$
for Reverse Longitude and so changes sign in the $u$-component of Eq. \ref{eq:NS}. Equivalently for Reverse Latitude,
$R_yR_y (-\tfrac{1}{\rho} \nabla p) =
-\tfrac{1}{\overset{\downarrow}{\rho}}(\partial_x \overset{\downarrow}{p}, -\partial_y \overset{\downarrow}{p})$
the sign changes in the $v$-component.

So overall, we have invariance for all terms except advection, Coriolis and pressure gradient that introduce some sign changes.
To counteract the sign change in the pressure gradient we negate the $u$-component of Eq. \ref{eq:NS}
for Reverse Longitude which yields for the advection term
\begin{equation}
R_x\m{~~\overset{\leftarrow}{u}\partial_x \overset{\leftarrow}{u} - \overset{\leftarrow}{v}\partial_y \overset{\leftarrow}{u} \\
      -\overset{\leftarrow}{u}\partial_x \overset{\leftarrow}{v} + \overset{\leftarrow}{v}\partial_y \overset{\leftarrow}{v}} =
R_x\m{(-\overset{\leftarrow}{u})\partial_x (-\overset{\leftarrow}{u}) + \overset{\leftarrow}{v}\partial_y (-\overset{\leftarrow}{u}) \\
      (-\overset{\leftarrow}{u})\partial_x \overset{\leftarrow}{v} + \overset{\leftarrow}{v}\partial_y \overset{\leftarrow}{v}} =:
R_x (\mathbf{\tilde{u}} \cdot \nabla) \mathbf{\tilde{u}}
\end{equation}

We therefore regain the conventional advection term when we solve for
$\mathbf{\tilde{u}} = (\tilde{u}, \tilde{v}) = (-\overset{\leftarrow}{u}, \overset{\leftarrow}{v})$ instead of $\mathbf{u}$,
meaning that reversing the initial conditions $u, v$ in longitude also requires a sign change in $u$.
Similarly, we have $\mathbf{\tilde{u}} = (\overset{\downarrow}{u}, -\overset{\downarrow}{v})$ for Reverse Latitude
such that here a sign change of $v$ becomes necessary. For the Coriolis term the negation of the $u$-component
of Eq. \ref{eq:NS} yields $R_x (-f \overset{\leftarrow}{v}, -f \overset{\leftarrow}{u})$ such that with
$\tilde{f} = -f$ we have $R_x (\tilde{f} \tilde{v}, -\tilde{f} \tilde{u})$ so that the conventional Coriolis term is regained
when using $\mathbf{\tilde{u}}$ and $\tilde{f} = -f$ which physically corresponds to the planet
rotating in the opposite direction. For Reverse Latitude the negation of the $v$-component of Eq. \ref{eq:NS}
yields $R_y (-f \overset{\downarrow}{v}, -f \overset{\downarrow}{u}) = R_y (f \tilde{v}, -f \tilde{u})$
so the conventional Coriolis term is regained when solving for $\mathbf{\tilde{u}}$ and
no change in the planetary rotation is necessary, $\tilde{f} = f$ here. In summary, the Navier-Stokes equations
are spatially invariant under reversal operators $R$
\begin{equation}
    R \left( \partial_t \mathbf{\tilde{u}} + (\mathbf{\tilde{u}} \cdot \nabla) \mathbf{\tilde{u}} + \tilde{f}\mathbf{k}
    \times \mathbf{\tilde{u}} = -\frac{1}{\tilde{\rho}}\nabla \tilde{p} + \nu \nabla^2\mathbf{\tilde{u}} + \mathbf{\tilde{F}} \right).
    \label{eq:NS_reversed}
\end{equation}
when solving for $\mathbf{\tilde{u}}, \tilde{f}, \tilde{\rho}, \tilde{p}, \mathbf{\tilde{F}}$ instead of $\mathbf{u}, f, \rho, p, \mathbf{F}$
with the tilde transforms being for Reverse Longitude
\begin{align}
\mathbf{\tilde{u}} = (-\overset{\leftarrow}{u}, \overset{\leftarrow}{v}), \quad
\tilde{f} = -f, \quad
\tilde{\rho} = \overset{\leftarrow}{\rho}, \quad
\tilde{p} = \overset{\leftarrow}{p}, \quad
\mathbf{\tilde{F}} = (-\overset{\leftarrow}{F_x}, \overset{\leftarrow}{F_y})
\end{align}
and for Reverse Latitude 
\begin{align}
\mathbf{\tilde{u}} = (\overset{\downarrow}{u}, -\overset{\downarrow}{v}), \quad
\tilde{f} = f, \quad
\tilde{\rho} = \overset{\downarrow}{\rho}, \quad
\tilde{p} = \overset{\downarrow}{p}, \quad
\mathbf{\tilde{F}} = (\overset{\downarrow}{F_x}, -\overset{\downarrow}{F_y}).
\end{align}

\subsection{Rotate Longitude}

For Rotate Longitude $R_o$ the sign changes due to the anticommutativity of $R_x, R_y$ with $\partial_x, \partial_y$
do not occur as we have $R_o\partial_x = \partial_xR_o$ and $R_o\partial_y = \partial_yR_o$ being commutative.
As we also have $R_of = f$ (Coriolis parameter is a function of latitude only) the Navier-Stokes equations are
invariant under longitudinal rotation when solving in this case for
\begin{align}
\mathbf{\tilde{u}} = (\overset{\circlearrowleft}{u}, \overset{\circlearrowleft}{v}), \quad
\tilde{f} = f, \quad
\tilde{\rho} = \overset{\circlearrowleft}{\rho}, \quad
\tilde{p} = \overset{\circlearrowleft}{p}, \quad
\mathbf{\tilde{F}} = (\overset{\circlearrowleft}{F_x}, \overset{\circlearrowleft}{F_y})
\end{align}
the rotated initial conditions and forcing without any further changes.

\subsection{Vorticity and divergence}

For models solving the equations of motions using vorticity $\zeta = \nabla \times \mathbf{u}$ and
divergence $D = \nabla \cdot \mathbf{u}$ instead of $u, v$ we have with Reverse Longitude
\begin{equation}
\zeta = R_x R_x (\nabla \times \mathbf{u}) =
R_x( -\partial_x \overset{\leftarrow}{v} - \partial_y\overset{\leftarrow}{u}) =
R_x(-\nabla \times (-\overset{\leftarrow}{u}, \overset{\leftarrow}{v}))
\end{equation}
which we identify as negative curl of the transformed velocity $\mathbf{\tilde{u}}$,
consequently called $\tilde{\zeta} = -\nabla \times \mathbf{\tilde{u}}$. The same sign change arises
with Reverse Latitude (but not for Rotate Longitude due to commutativity) such that also there
one has to solve for the negated (and reversed) vorticity $\tilde{\zeta} = -R\zeta$ instead of $\zeta$.
For divergence $D$ we have
\begin{equation}
D = R_x R_x (\nabla \cdot \mathbf{u}) =
R_x( -\partial_x \overset{\leftarrow}{u} + \partial_y\overset{\leftarrow}{v}) =
R_x(\nabla \cdot (-\overset{\leftarrow}{u}, \overset{\leftarrow}{v})).
\end{equation}

As no sign change occurs, $\tilde{D} = \nabla \cdot \mathbf{\tilde{u}}$ and so the equations
are solved for $\tilde{D} = RD$.

\section{Supplementary figures}

\begin{figure}[H]
\noindent\includegraphics[width=\textwidth]{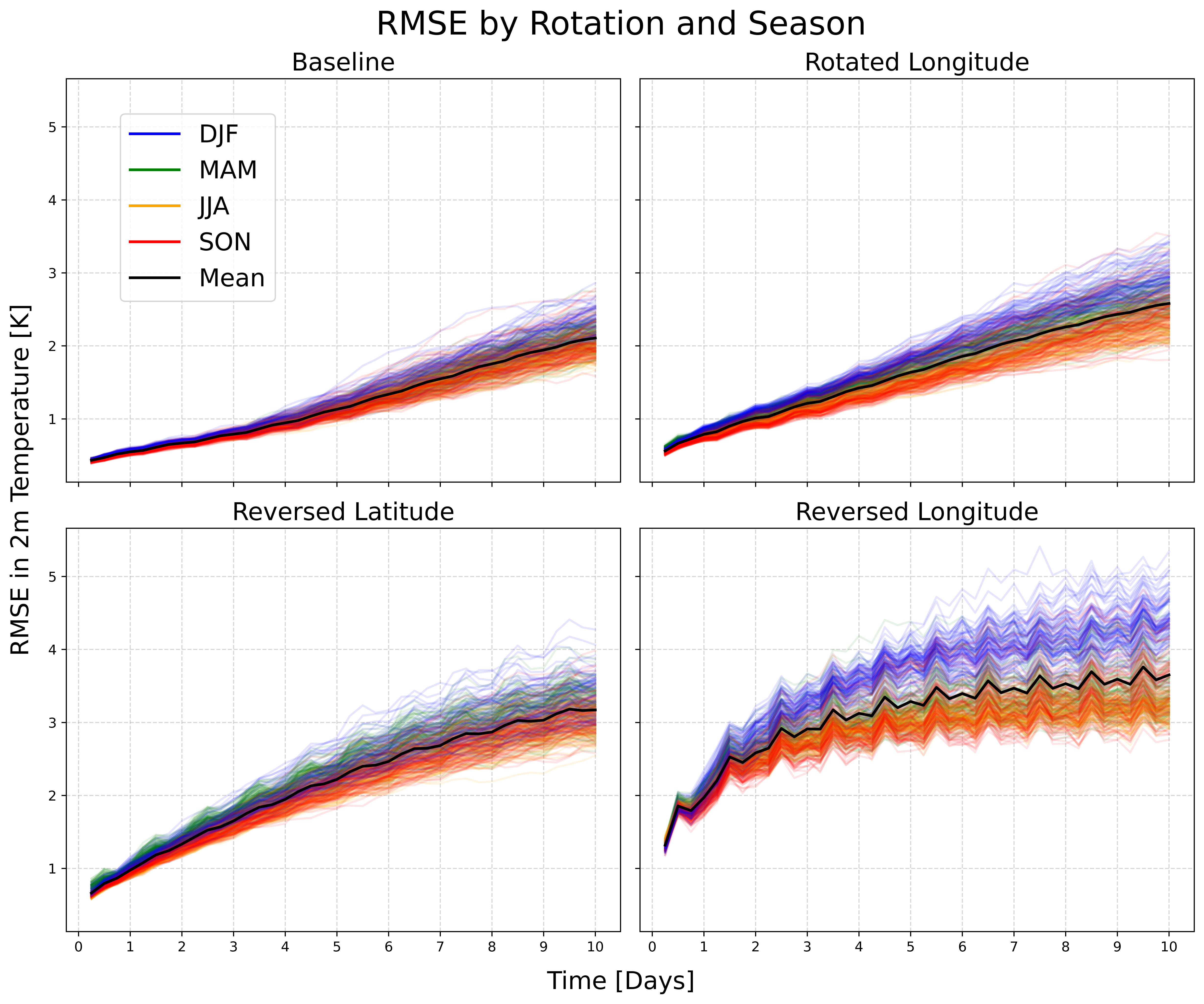}
    \caption{GraphCast generalization error as Fig.~\ref{fig:gen_error} but by season (DJF = December, January, February, etc.) in 2022.}
    \label{fig:by_seasons}
\end{figure}

\begin{figure}[H]
\noindent\includegraphics[width=\textwidth]{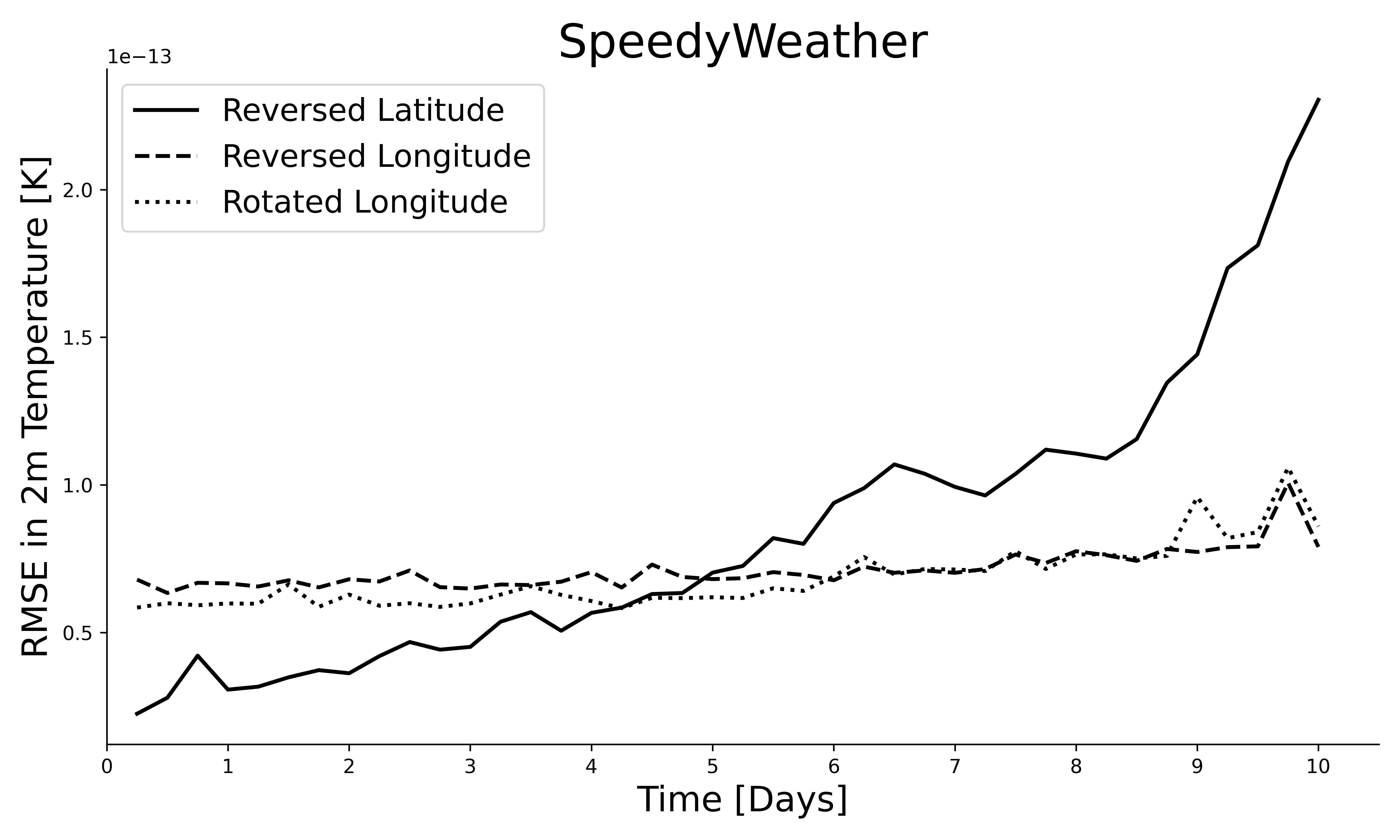}
    \caption{\textbf{SpeedyWeather's generalization error by test case}.
    Note that errors are in order of magnitude $10^{-13}$. 
    Simulations are computed using 64-bit double-precision floating-point numbers and errors scale with precision,
    i.e., single-precision errors are several orders of magnitude higher.
    Generalization error computed as described in Section~\ref{error_calc}.
}
    \label{fig:SpeedyErrors}
\end{figure}

\begin{figure}[H]
\noindent\includegraphics[width=\textwidth]{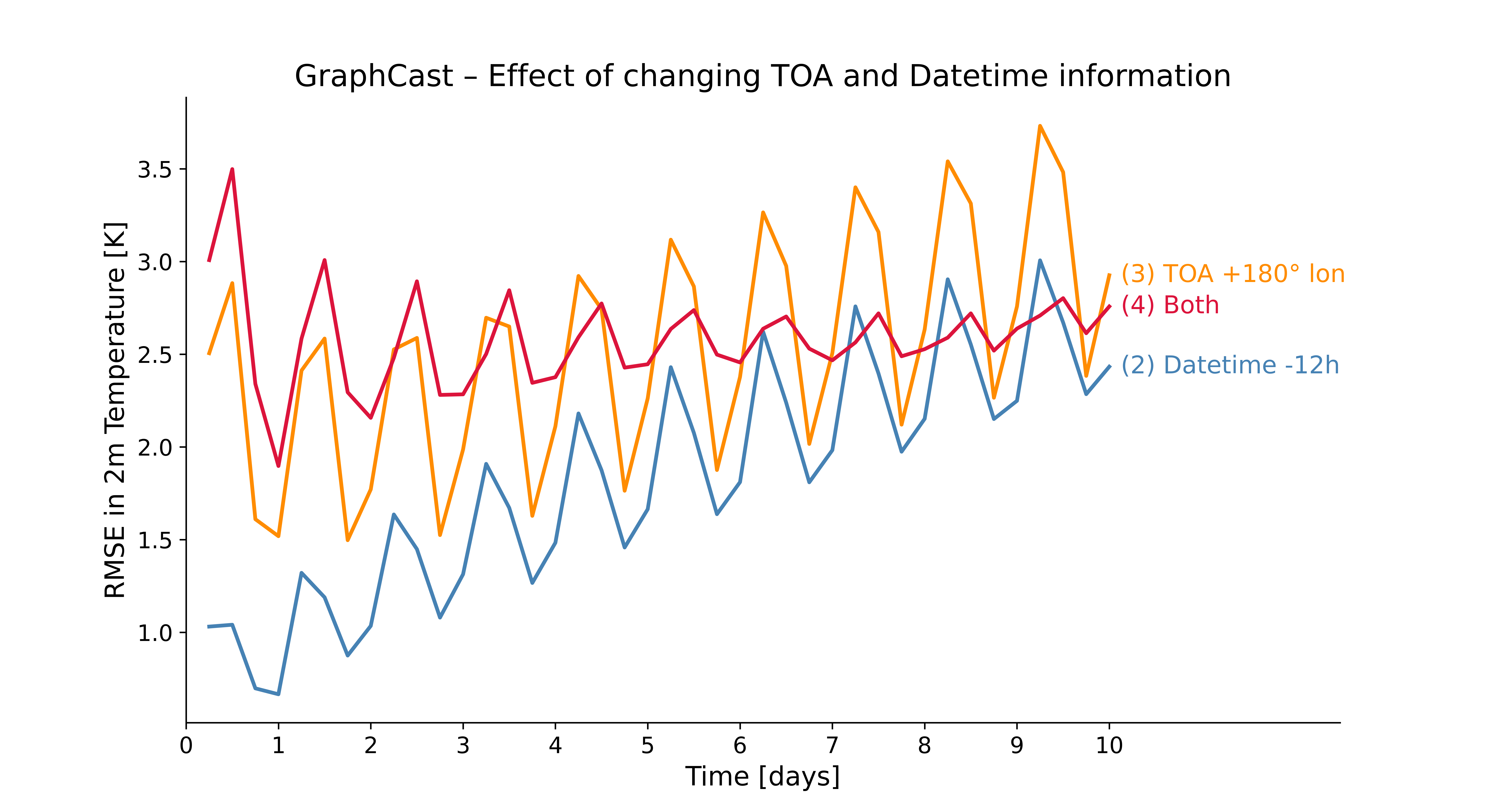}
    \caption{\textbf{RMSE in 2m-temperature for different daily cycle adaptations.}
        Same as Fig.~\ref{fig:dependence_on_TOA_and_datetime}, but spatially averaged. Area-weighted RMSE in 2m-temperature introduced by shifting date-time by -12h (blue), rotating top-of-atmosphere incoming solar radiation by 180\textdegree{} in longitude (orange), and performing both adaptations (red) compared to a baseline forecast.
        All 10-day forecasts were initialized on 2022-01-01 with a 6h timestep.}
    \label{fig:dependence_on_TOA_and_datetime_average}
\end{figure}

\begin{figure}[H]
\noindent\includegraphics[width=\textwidth]{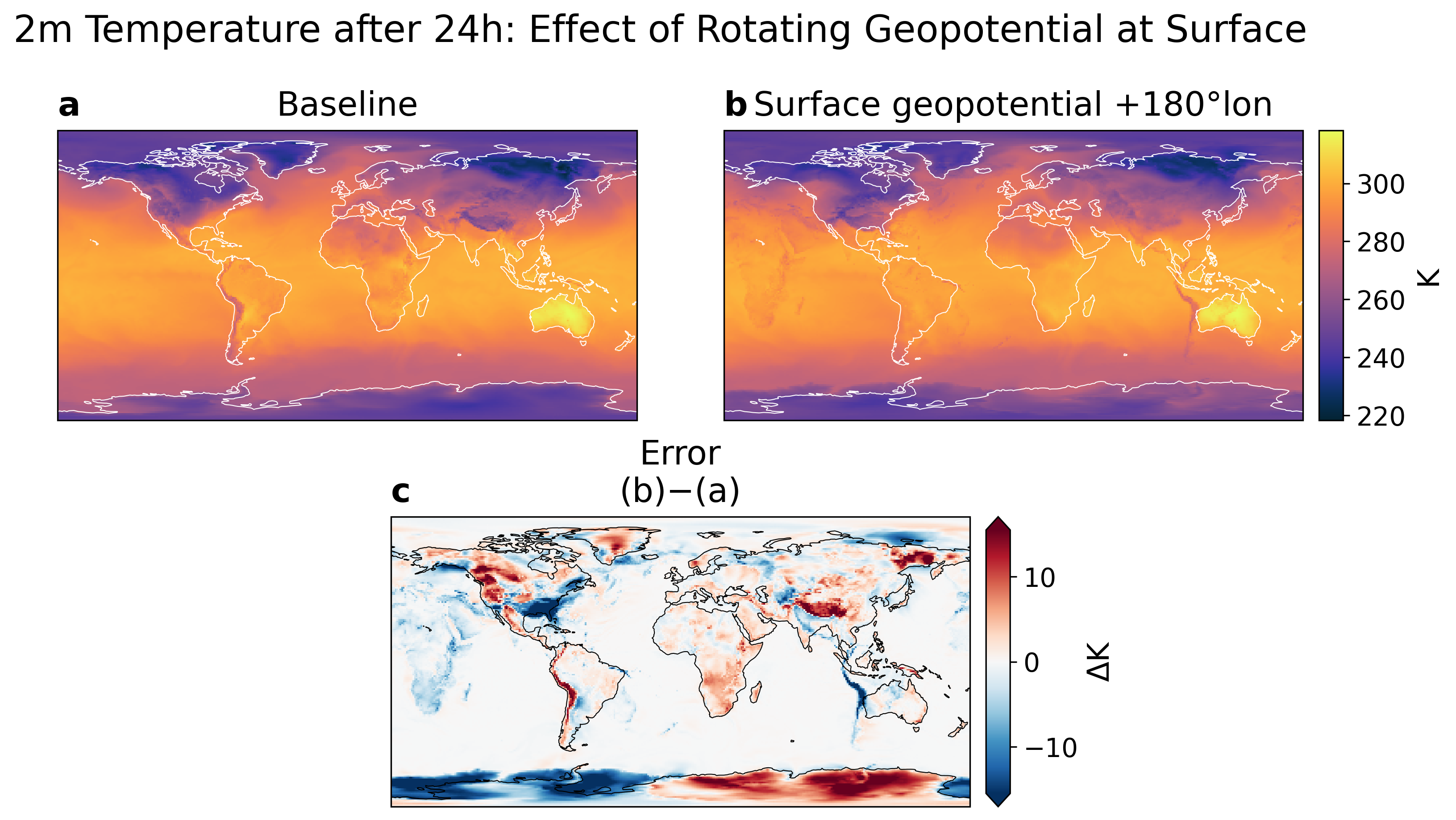}
    \caption{\textbf{Effect of changing geopotential at surface on 2m-temperature.} (a) GraphCast 24-hour forecast of 2m-temperature initialized on 2022-01-01, (b) as (a) but geopotential at surface is shifted by 180\textdegree{} in longitude, (c) difference between (a) and (b).}
    \label{fig:dependence_on_geopot_and_lon}
\end{figure}

\section*{Data availability}
The ensembles of GraphCast forecasts used to produce the figures in this paper can be found on Zenodo, which will be provided as a DOI upon acceptance. The ERA5 reanalysis data used to initialize the forecasts is available as a Google Cloud Public Dataset~\citep{ARCOERA5AnalysisReadyCloudOptimized, hersbachERA5GlobalReanalysis2020}. Specifically, we use \nolinkurl{'gs://gcp-public-data-arco-era5/ar/full_37-1h-0p25deg-chunk-1.zarr-v3'} and re-grid to the 1$\textdegree{}$ resolution of the model as shown in the code.

\section*{Code availability}
Code to reproduce results is published here \url{https://github.com/marenhoever/Spatial_Generalization_Tests.git} (will be converted to DOI upon acceptance).
 Data is processed using Xarray~\citep{hoyerXarrayNDLabeled2017} and NumPy~\citep{harrisArrayProgrammingNumPy2020}. Figures were made with Matplotlib~\citep{hunterMatplotlib2DGraphics2007} and Cartopy~\citep{Cartopy}. Fig.~\ref{fig:planets} was created using GeoMakie.jl~\citep{MakieOrgGeoMakiejl2026}. Simulations were performed on the JASMIN computing cluster~\cite{lawrenceJASMINSuperdatacluster2012, JASMINSite}.

\section*{Acknowledgments}
We are grateful to the GraphCast and NeuralGCM teams for insightful discussions on this project.
MH would like to thank David Marshall and Andrew Wells for their constructive feedback on a previous version of this manuscript.
This work was conducted as part of the Intelligent Earth CDT supported by funding from the UK Research and Innovation Council (UKRI) grant number EP/Y030907/1.
MH acknowledges funding by the Rhodes Trust. MK acknowledges funding from the Natural Environment Research Council under grant number UKRI191. CSDW acknowledges funding from the Department for Science, Innovation and Technology and the Royal Academy of Engineering under the Research Fellowships scheme. HMC was supported by a Leverhulme Trust Research Leadership Award `Seamless Uncertainty Quantification for Earth System prediction' (SUQCES).

\section*{Author Contributions}
MK conceptualized the original test cases. 
They were further developed and implemented by MH, MK, and HMC.
MH conducted the experiments and did the evaluation.
MK, HMC, and CSDW supervised the work and gave feedback throughout.
MH wrote the first draft of the paper, MK wrote sections~\ref{sec:math_background},~\ref{sec:grid_explainations},~\ref{sec:grid_explainations_spherical}, and appendix~\ref{proof}.
The paper was then reviewed and edited by MK, HMC, MH, and CSDW.

\section*{AI use statement}
For the work presented in this article, ChatGPT Edu, Gemini, and Claude Code supported coding at different stages throughout the project.
ChatGPT Edu was used to probe reasoning and check for errors during the conception of the project.
The writing is solely the authors'.
However, ChatGPT Edu has been used to find better logical ordering of arguments, as well as for high-level suggestions or slight reformulations.

\bibliography{references, milans_references}
\end{document}